\documentclass[aps,prd,showpacs,twocolumn,amsmath]{revtex4}

\usepackage{graphicx}
\usepackage{bbm}

\newcommand{\be}{\begin{eqnarray}}
\newcommand{\ee}{\end{eqnarray}}
\newcommand{\nn}{\nonumber\\}

\newcommand{\sfrac}[2]{{\textstyle\frac{#1}{#2}}}

\newcommand{\m}{\mathbf{m}}
\newcommand{\F}{\mathcal{W}}
\newcommand{\tr}{\mathrm{tr}}

\begin{document}

\title{Linear confinement without dilaton in bottom-up holography for walking technicolour}

\author{Dennis D.~Dietrich}
\author{Matti J\"arvinen}
\affiliation{CP$^3$-Origins, University of Southern Denmark, Campusvej 55, DK-5230 Odense M, Denmark}

\author{Chris Kouvaris}
\affiliation{The Niels Bohr Institute, Blegdamsvej 17, DK-2100 Copenhagen {\O}, Denmark}

\date{September 1, 2009}

\begin{abstract}

In~\cite{Dietrich:2008ni,Dietrich:2008up}, we constructed a holographic description of walking technicolour theories 
using both a hard- and a soft-wall model. Here, we show that 
the dilaton field becomes phenomenologically irrelevant for the spectrum of spin-one resonances once a term is included
in the Lagrangian that mixes the Goldstone bosons and the longitudinal components of the axial vector mesons. We show how this mixing affects our previous results and we make predictions about how this description of technicolour can be tested.

\end{abstract}


\maketitle


\section{Introduction}

Since walking, i.e., quasi-conformal technicolour~\cite{TC} theories were identified, which are consistent with available data~\cite{Sannino:2004qp,Dietrich:2006cm},
they have been in the centre of intensive study. 
Theories of this type, do not only break the electroweak symmetry naturally, but in addition, they can lead to possible new extra phase transitions~\cite{Cline:2008hr} and provide several dark matter candidates that avoid experimental constraints~\cite{Gudnason:2006yj}.

Generally speaking, it is a very difficult task to carry out computations in strongly interacting theories like technicolour, due to the failure of the perturbative expansion. 
Already the identification of candidates for walking technicolour theories presupposes knowledge about the conformal window of strongly interacting theories \cite{Dietrich:2006cm,Ryttov:2007sr}. 
For this reason, different approaches have to be implemented. They include
effective theories, lattice studies and holography. The first exploration of some of the aforementioned technicolour models on the lattice~\cite{Catterall:2007yx} confirm that they do not show running behaviour. Also effective theories have been constructed for a selection of models~\cite{Gudnason:2006ug,Foadi:2007ue}, to make predictions for collider phenomenology for the LHC.

Although effective theories can provide a useful tool in the study of strongly coupled
theories, they usually suffer from a large number of unknown parameters that can be fixed only when compared against experimental results. From this point of view, holographic
descriptions of quasi-conformal theories might be able to provide more stringent constraints than the ``usual'' four dimensional effective theory. Holographic descriptions of QCD~\cite{Erlich:2005qh,Karch:2006pv,Da Rold:2005zs,Brodsky:2008pg} turned out to be in good agreement with the experimental results. 
Therefore, if one accepts holography~\cite{Maldacena:1997re}
as a legitimate tool to probe strongly coupled theories, it is expected that, since by definition walking technicolour theories~\cite{walk} are
quasi-conformal (in a range of energy), holography should describe them better than theories of the QCD type [which are far from being (quasi)-conformal]~\cite{Hong:2006si,Dietrich:2008ni}. Extra-dimensional holographic frameworks have also been employed to construct technicolour-like models
\cite{Hirn:2006nt}.

In~\cite{Dietrich:2008ni,Dietrich:2008up}, we studied a holographic implementation of walking technicolour theories. Our motivation was to constrain the parameters of the effective Lagrangian and make predictions for the mass spectrum of the theories. This was done by implementing two different setups for holography, i.e., the hard-~\cite{Erlich:2005qh,Polchinski:2000uf} and the soft-wall~\cite{Karch:2006pv} models. In the hard-wall model, we assumed the technicolour theory under investigation to be conformal between two energy scales, namely, an infrared one, which coincides with the electroweak scale, and an ultraviolet scale beyond which the coupling runs again. 
The soft-wall model has a smooth cut off. There, the IR scale is set by the inverse slope of a dilaton background. This prescription can produce the proper scaling of the spectrum in QCD~\cite{Karch:2006pv}.

In~\cite{Dietrich:2008ni}, we examined the constraints arising for a general walking technicolour theory, where the vector mesons do not couple directly to the condensate. In this particular setup, using the aforementioned holographic prescription, we were able to find a relation between the mass of the axial and the vector mesons, as well as their decay constants. In this case, the mass of the axial was always larger than that of the vector mesons, which was bounded from below. In~\cite{Dietrich:2008up}, we generalised our investigation, by allowing terms in our effective theory that could couple the vector mesons to the chiral condensate. This was done by including terms admitted by the global symmetry, and then promoting the global symmetry to a local one, using a St\"uckelberg completion~\cite{DDD1}.
In this case, there is one additional free parameter, the one parametrising the coupling of the vector mesons to the condensate. In the generalised picture, the mass hierarchy between vector and axial vector mesons can be inverted, although the lower bound for the axial mass persists. 
From this treatment we left out a current term because it does not affect the equations of motion in the bulk for the vector and axial mesons. As it turns out, however, it does alter the decay constants. As we shall argue in this paper, the existence of this term allows for the possibility of mixing between the Goldstone bosons and the longitudinal component of the axial vectors, leading to the change in the decay constant of the axial vectors. (In fact, theoretically speaking, the decay constant can even be zero, due to deconstructive  interference between the Goldstone bosons and the axial vectors.) We shall show that in the presence of said current term, there is no need for a dilaton field, and effectively, we can omit the dilaton.
The presence of the current term enlarges the parameter space even after we have removed the dilaton. It turns out that the lowest bound on the mass of the lighter (axial)-vector meson remains even after the inclusion of the aforementioned mixing, and it is a robust prediction of our model. 

Our paper is organised as follows: In section II, we review our notation and our formalism and we present all the relevant terms for the effective theory that we use. In section III, we show how the mixing between the Goldstone bosons and the longitudinal component of the axial vector arises, and how it affects our analysis. In section IV, we compare various holographic setups. In section V, we summarise our results.



\section{Framework\label{II}}

Regard chiral symmetry breaking from a global flavour symmetry group $\mathcal{G}$ to $\mathcal{H}$.
For technicolour with $N_f$ techniquarks in non-(pseudo)real representations of the technicolour gauge group, we would have $\mathcal{G}=SU(N_f)_L\times SU(N_f)_R$ and $\mathcal{H}=SU(N_f)_V$.
In the case of real [pseudoreal] representations, there are enhanced breaking patterns, $\mathcal{G}=SU(2N_f)$ and $\mathcal{H}=SO(2N_f)$ [$\mathcal{H}=Sp(2N_f)$], always assuming the breaking to the diagonal subgroup.

Define the fermion bilinear (with contraction of the spin indices),
$
M=\epsilon_{\alpha\beta}Q^\alpha\otimes Q^\beta ,
$
where
\be
Q=(U_L,D_L,\dots,-i\sigma^2U_R^*,-i\sigma^2D_R^*,\dots)^T .
\ee
Here, $U_{L/R}$, $D_{L/R}$ are the left-/right-handed fermion fields. [For non-(pseudo)real representations the components of $M$ involving left- and right-fields simultaneously are technicolour non-singlets and are discarded.] The chiral condensate be proportional to the matrix $E$. The $S^a$ with $ES^aE=-S^{aT}$ span $\mathcal{H}$. The remaining generators $X^a$ of $\mathcal{G}$ are the generators for the pion field.
The spin-one degrees of freedom $A_\mu=A^a_\mu T^a$
span the entire ${\cal G}$.
With these fields we formulate the Lagrangian ${\cal L}$. The part depending on $A_\mu^a$ reads,
\begin{equation}
{\cal L}
=
-\sfrac{1}{4g^2_5}F^a_{\mu\nu}F^a_{\kappa\lambda}g^{\mu\kappa}g^{\nu\lambda}
+\sfrac{1}{2} A^a_\mu\m^{ab}A^b_\nu g^{\mu\nu}
+A_\mu^a J^a_\nu g^{\mu\nu} ,
\nonumber
\end{equation}
where $F^a_{\mu\nu}$ stands for the field tensor $F^a_{\mu\nu}T^a=F_{\mu\nu}=\partial_\mu
A_\nu-\partial_\nu A_\mu-i[A_\mu,A_\nu]$. $\m^{ab}=\m^{ba}$ and $J^a_\mu$ depend on $M$ in such a way that $\mathcal{L}$ is invariant under simultaneous {\it global} transformations, $U\in{\cal G}$,
$A_\mu \rightarrow UA_\mu U^\dagger$
and
$M \rightarrow UMU^T$ .
Hence, the allowed expressions are,
\be
\sfrac{1}{2}A^a_\mu\m^{ab}A^b_\nu g^{\mu\nu}
&=&
[r_1~\tr(A_\mu A_\nu MM^\dagger)
+\label{mass}\\
&&+
r_2~\tr(A_\mu M A_\nu^T M^\dagger)
+
\nn
&&+
s~\tr(A_\mu A_\nu)\tr(MM^\dagger)/(2N_f)]g^{\mu\nu} ,
\nonumber
\ee
\begin{equation}
J^a_\mu
=
ir_3\tr
\{T^a[M(\partial_\mu M)^\dagger-(\partial_\mu M)M^\dagger]\} .
\label{curr}
\end{equation}

In holography in the bulk, the flavour symmetry is treated as a local symmetry. Hence, the action is
to be invariant under local inhomogeneous transformations of the spin-one fields,
$
A^a_\mu T^a=A_\mu\rightarrow U[A_\mu-iU^\dagger(\partial_\mu U)]U^\dagger,
$
accompanied by local homogeneous transformations of $M$.
They do not leave invariant the mass (\ref{mass}) and the current term (\ref{curr}), unless they combine into $\sim\tr[(D_\mu M)(D_\nu M)^\dagger]g^{\mu\nu}$, where $D_\mu M:=\partial_\mu M-iA_\mu M-iMA_\mu^T$. In general, the invariance under inhomogeneous transformations can be ascertained by introducing non-Abelian St\"uckelberg degrees of freedom $\Phi=e^{-i\theta^a T^a}$, which live in a copy $\mathcal{G}^\prime$ of the symmetry group
$\mathcal{G}$, where $\Phi\rightarrow U\Phi$. Replacing all spin-one fields according to
$
A_\mu\mapsto A^\Phi_\mu=A_\mu-i\Phi(\partial_\mu\Phi)^\dagger
$
and all partial derivatives by the covariant derivative,
\begin{equation}
\partial_\mu M\mapsto D^\phi_\mu M
=\partial_\mu M+\Phi(\partial_\mu\Phi)M+M[\Phi(\partial_\mu\Phi)]^T ,
\label{repl2}
\end{equation}
ensures the gauge invariance Eqs.~(\ref{mass}) and (\ref{curr}), as $A_\mu^\Phi\rightarrow
UA^\Phi_\mu U^\dagger$ and $D^\Phi_\mu M\rightarrow U(D_\mu^\Phi M)U^T$. \{This corresponds to the introduction of the field $N$ in Ref.~\cite{Foadi:2007ue}. (See also Appendix A of \cite{Dietrich:2008up}.)\}
The original symmetry content is restored after $\Phi$ takes a constant vacuum expectation value. As $\Phi$ only appears in combination with derivatives, the expectation value does not contribute.

In the transition from a global to a local symmetry, also initially present global anomalies have to be accounted for. According to the reasoning employed in the derivation of the 't Hooft anomaly matching conditions, anomalies in bottom-up holography can be cancelled by introducing an
appropriate number of additional fermions, which are sufficiently weakly interacting
as not to affect the dynamics of the system.

Another way to avoid, e.g., the $SU(N_f)^2 U(1)$ anomalies
[$SU(N_f)$ can be the left- or right subgroups and $U(1)$ the technibaryon number, which is gauged in set ups in which the elementary fermions transform under (pseudo)real representations of the (colour/technicolour) gauge group]
is to gauge merely the $SU(N_f)_L\times SU(N_f)_R$ subgroup of $\mathcal{G}$ and not the full $SU(2N_f)$, which also contains the $U(1)$. The $SU(N_f)_L\times SU(N_f)_R$ subgroup is designated anyhow by the coupling to the electroweak interactions.

Taking stock, the part of the five-dimensional action in unitary gauge, which is relevant for the subsequent computations is given by,
\be
S:=\int d^5x \sqrt{g} ({\cal L}+{\cal L}_M),
\label{action}
\ee
and
\be
{\cal L}_M
:=
\tr[(\partial_\mu M)(\partial_\nu M)^\dagger] g^{\mu\nu}
-
m^2_5\tr(MM^\dagger).
\ee
The metric $g$ be anti de Sitter,
$
ds^2=z^{-2}(-dz^2+dx^2).
$
The fifth coordinate $z$ is interpreted as inverse energy scale.
Matching in the ultraviolet fixes the coupling $g_5$ to \cite{Erlich:2005qh}
\be
g^2_5=12\pi^2/d_\mathrm{R},
\label{uv}
\ee
where $d_R$ stands for the dimension of the technicolour gauge group with respect to which the fermions transform.

The soft-wall model \cite{Karch:2006pv} contains an additional dilaton field $\phi$,
\be
S_s:=\int d^5x\sqrt{g}e^{-\phi}(\mathcal{L}+\mathcal{L}_M).
\ee
For $\phi=cz^2$, $c=\mathrm{constant}$, the mass spectrum of the spin-one resonances features a linear spacing in the squared masses.

The condensate $M_0$ of $M$ obeys the differential equation,
\be
(z^3e^{+cz^2}\partial_z z^{-3}e^{-cz^2}\partial_z-z^{-2}m^2_5)
M_0
=0.
\label{eomm0}
\ee
For $c^2z^2\ll m^2_5$, i.e., in the ultraviolet, the ansatz $M_0\sim z^d$ leads to the
characteristic equation for the dimension $d$,
$
m^2_5=d(d-4).
$
In the quasi-conformal case $d=2$ and
this implies $m^2_5=-4$. Eq.~(\ref{eomm0}) is solved by
$
g_5M_0
=C z^2 e^{+cz^2}.
$
Hence, the condensate $M_0$ grows exponentially with $z$. This
is indicative of an instability, which, ultimately, must be caught by higher-order terms in the equation of motion \cite{Karch:2006pv}, originating from potential terms for the spin-zero fields in the Lagrangian, which have not been incorporated here. As the solution is thus adequate for small $z$, we set
\be
g_5M_0
\mapsto Cz^2.
\label{M0s}
\ee
This can be seen as a practical way of regularising said instability. In fact, there
are nonlinear terms which can alter Eq.~(\ref{eomm0}) such that Eq.~(\ref{M0s}) is an
exact solution. For the interpretation of the preceding step as $O(cz^2)$ approximation, the propagation of the error was assessed in \cite{Dietrich:2008up}.

After a Fourier transformation to four-dimensional momentum ($q$) space the equation of motion for $A_\perp^a$, the transverse part of the spin-one field, reads
\begin{equation}
[
(ze^{+cz^2}\partial_z z^{-1}e^{-cz^2}\partial_z+q^2)
\delta^{ab}-g^2_5z^{-2}\mathbf{m}^{ab}]A^b_\perp=0.
\end{equation}
By diagonalising in flavour space, we find
\begin{equation}
[
(ze^{+cz^2}\partial_z z^{-1}e^{-cz^2}\partial_z+q^2)
-g^2_5z^{-2}M_0^2\lambda^2_\F]\F
=0.
\label{spinone}
\end{equation}
Here and in what follows, $\F$ represents the vector
$\mathcal{V}$ and the axial vector $\mathcal{A}$ eigensolution,
respectively.
\begin{equation}
\lambda^2_\mathcal{V}=r_1-r_2+s\mathrm{~~~and~~~}
\lambda^2_\mathcal{A}=r_1+r_2+s
\end{equation}
stand for the eigenvalues of $\m^{ab}/M_0^2$.
For the ultraviolet boundary conditions and with $M_0$ from Eq.~(\ref{M0s}), Eq.~(\ref{spinone}) is solved by
\be
\F
\sim
z^2e^{-(c_\F-c) z^2/2}M(1-\sfrac{M_\F^2}{4c_\F},2,c_\F z^2),
\ee
where $c^2_\F:=c^2+C^2\lambda_\F^2$. The normalisation condition
\be
\int_0^\infty\frac{dz}{z}e^{-cz^2}\F^2=1,
\ee
can only be satisfied for $M^2_\F=4nc_\F$, $n\in\mathbbm{N}$, i.e., when the series for the Kummer function $M(a,b,x)$ truncates into a polynomial. For $n=1$, the normalised solution reads,
\be
\F=\sqrt{2}c_\F z^2e^{-(c_\F-c)z^2/2}.
\ee

For a short while, let us continue for the special case $r_3=0$. There
the decay constants can be extracted according to,
\be
g_5F_\F=\partial_z^2\F(0)=2\sqrt{2}c_\F=M_\F^2/\sqrt{2}.
\label{softdecay}
\ee

For determining the pion decay constant $f_\pi$ we have to solve,
\be
[ze^{+cz^2}\partial_z z^{-1}e^{-cz^2}\partial_z-g^2_5z^{-2}M_0^2\lambda^2_\mathcal{A}]\mathcal{P}
=0.
\ee
Taking $M_0$
from (\ref{M0s}) and for the boundary conditions
$
\partial_z\mathcal{P}(z_m)=0$ and $\mathcal{P}(0)=1,
$
this equation is solved by,
$
\mathcal{P}
=e^{-(c_\mathcal{A}-c)z^2/2}.
$
This leads to the pion decay constant
\be
g^2_5f_\pi^2
=
-\lim_{\epsilon\rightarrow 0}\partial_z\mathcal{P}(\epsilon)/\epsilon
\overset{\partial_z\mathcal{P}(0)=0}{=}
-\partial^2_z\mathcal{P}(0)
=
c_\mathcal{A}-c
.
\label{fpi}
\ee


\section{The effect of the  ${\bf r_3}$ term\label{r3term}}

While the $r_3$ term does not have an impact on the above equations of motion, as it vanishes when evaluated on the expectation value of $M$, it changes the coupling of the pion and axial vector fields to the 
electroweak sector
and hence the physical values of the decay constants. Let us first work out explicitly how the couplings are modified in four dimensional effective theory. 
For simplicity we use an effective Lagrangian that describes the $SU(2)_{\rm L}\times SU(2)_{\rm R} \to SU(2)_{\rm V}$ chiral symmetry breaking pattern rather than the
potentially enhanced pattern in other settings like the
$SU(4)\to SO(4)$ breaking pattern of Minimal Walking Technicolour (MWT). The used Lagrangian is obtained from
those involving more fields due to an enlarged symmetry
by removing fields that would not enter the derivation.

Following \cite{Foadi:2007ue}, we use the hidden local symmetry approach, where the composite vector fields are introduced as gauge fields of a copy $SU(2)_{\rm L}'\times SU(2)_{\rm R}'$ of the chiral symmetry group.
The nonlinear sigma fields which connect this copy to the actual chiral group read
\begin{eqnarray}
N_{\rm L}=\exp\left(\,2 i \widetilde{\pi}^a_{\rm L} T^a / f\right) \ , \quad
N_{\rm R}=\exp\left(\,2 i \widetilde{\pi}^a_{\rm R} T^a / f\right)
\end{eqnarray}
and the linear sigma field that contains the (pseudo) scalar degrees of freedom is written as
\begin{eqnarray}
M=\frac{1}{\sqrt{2}}\left(v+H-2 i \widetilde{\pi}^a T^a\right) \ .
\end{eqnarray}
Here, $H$ is the composite Higgs, the normalisation of the $SU(2)$ generators is ${\rm Tr}\left[T^aT^b\right]=\delta^{ab}$ and the pion fields
\begin{equation}
\widetilde{\pi}_{\rm A} = \widetilde{\pi}_{\rm L}+\widetilde{\pi}_{\rm R}  \ , \quad
\widetilde{\pi}_{\rm V} = \widetilde{\pi}_{\rm L}-\widetilde{\pi}_{\rm R}
\end{equation}
will be eaten by the composite vector bosons when $N_{\rm L/R}$ receive their vacuum expectation values.

The relevant terms in the Lagrangian which affect the quadratic vector sector are
\be
\mathcal{L} &=& \frac{f^2 }{4}{\rm Tr}[(D_\mu N_{\rm L})^\dagger (D^\mu N_{\rm L}) + (D_\mu N_{\rm R})^\dagger (D^\mu N_{\rm R})] \nonumber \\
&&+\frac{1}{2}{\rm Tr}[(\partial_\mu M)^\dagger (\partial^\mu M)] \nonumber \\
&&+ r_2{\rm Tr}[(D_\mu N_{\rm L})^\dagger N_{\rm L} M (D^\mu N_{\rm R}) N_{\rm R}^\dagger M^\dagger] \nonumber \\
&&- \frac{r_3}{2
}{\rm Tr}\{(D_\mu N_{\rm L})^\dagger N_{\rm L}[M (\partial^\mu M)^\dagger-(\partial^\mu M) M^\dagger] \nonumber \\
&& +  (D^\mu N_{\rm R}) N_{\rm R}^\dagger [M^\dagger (\partial_\mu M) - (\partial_\mu M)^\dagger M]\} ,
\ee
where the covariant derivatives read
\be
D_\mu N_{\rm L}&=&\partial_\mu N_{\rm L}-i\ \tilde{g}\ A_{{\rm L}\mu}^a\ T^a\ N_{\rm L}
\nonumber \\
D_\mu N_{\rm R}&=&\partial_\mu N_{\rm R}
+i\ \tilde{g}\ N_{\rm R}\ A_{{\rm R}\mu}^a\ T^a
\ee
with the composite vectors
\begin{equation}
A_\mu = \frac{1}{\sqrt{2}}\left(A_{{\rm L}\mu} - A_{{\rm R}\mu}\right) \ , \quad V_\mu = \frac{1}{\sqrt{2}}\left(A_{{\rm L}\mu} + A_{{\rm R}\mu}\right) \ .
\end{equation}
We excluded the $r_1$ and $s$ terms discussed above, since their effect on the quadratic Lagrangian can be absorbed into the above terms.
The vector and axial masses in terms of the parameters of the Lagrangian are
\begin{equation}
M_{\rm A}^2 = \frac{\tilde g^2}{4}\left(f^2+r_2v^2\right) \ , \quad M_{\rm V}^2 = \frac{\tilde g^2}{4}\left(f^2-r_2v^2\right) \ .
\end{equation}

We shall derive the decay constants by studying the mixing of the pions. The vector-pion mixing term reads
\be
 \mathcal{L}_{\rm V-\pi} &=& -\tilde g
\left(\begin{array}{c}
A^\mu \\
 V^\mu
\end{array}\right)^T \mathcal{C}_{\rm mix}
\left(\begin{array}{c}
 \partial_\mu \widetilde \pi \\
 \partial_\mu \widetilde \pi_{\rm A} \\
 \partial_\mu \widetilde \pi_{\rm V}
\end{array}\right) ,
\ee
where
\be
  \mathcal{C}_{\rm mix} &=& \left(\begin{array}{ccc}
 -\frac{r_3 v}{
 \sqrt{2}} & \frac{f^2+r_2 v^2}{2 \sqrt{2}f} & 0 \\
 0& 0 & \frac{f^2-r_2 v^2}{2\sqrt{2} f}
\end{array}\right) \ .
\ee
It can be used to identify the pions $\pi_{\rm A/V}$ that are eaten by the composite vectors. We see that these are not exactly the pions $\widetilde \pi_{\rm A/V}$, but there is some mixing between the pion $\widetilde \pi$ and the axial vector because of the $r_3$ term:
\begin{equation}
 \left(\begin{array}{c}
M_{\rm A} \pi_{\rm A} \\
 M_{\rm V} \pi_{\rm V}
\end{array}\right) =
\tilde g \ \mathcal{C}_{\rm mix}\left(\begin{array}{c}
 \widetilde \pi \\
 \widetilde \pi_{\rm A} \\
 \widetilde \pi_{\rm V}
\end{array}\right)  =
\left(\begin{array}{c}
\frac{\sqrt{2} M_{\rm A}^2}{\tilde g f}  \widetilde \pi_{\rm A} - \frac{\tilde g r_3 v}{
\sqrt{2}}\widetilde \pi \\
 \frac{\sqrt{2} M_{\rm V}^2}{\tilde g f}  \widetilde \pi_{\rm V}
\end{array}\right) \ .
\end{equation}
By inserting this back to the Lagrangian, the pion kinetic term becomes
\begin{equation}
\mathcal{L}_{\rm \pi} = \frac{1}{2}\!\left[\left( \partial_\mu \pi_{\rm A}\right)^2\! + \!\left( \partial_\mu \pi_{\rm V}\right)^2 \right.
\!+\! \left. \left(1\!-\!\frac{\tilde g^2r_3^2v^2}{2
M_{\rm A}^2} \right)\left(\partial_\mu \widetilde \pi\right)^2\right] .
\nonumber
\end{equation}
Hence, the 
normalisation of the physical pion must be adjusted 
to compensate for the part eaten by the axial vector. Taking this into account, we have
\be
 \left(\begin{array}{c}
  \pi \\
  \pi_{\rm A} \\
  \pi_{\rm V}
\end{array}\right) & = &
\left(\begin{array}{ccc}
  \sqrt{1\!-\!\frac{\tilde g^2r_3^2v^2}{2
  M_{\rm A}^2}} & 0 & 0 \\
  -\frac{\tilde g r_3 v}{
  \sqrt{2} M_{\rm A}} & \frac{\sqrt{2} M_{\rm A}}{\tilde g f} & 0 \\
  0 & 0 &  \frac{\sqrt{2} M_{\rm V}}{\tilde g f}
\end{array}\right)\left(\begin{array}{c}
 \widetilde \pi \\
 \widetilde \pi_{\rm A} \\
 \widetilde \pi_{\rm V}
\end{array}\right) \nonumber \\
&\equiv& \mathcal{C}_\pi
\left(\begin{array}{c}
 \widetilde \pi \\
 \widetilde \pi_{\rm A} \\
 \widetilde \pi_{\rm V}
\end{array}\right) \ .
\ee

Then the decay constants are related to the vacuum expectation values $v,f$ by the matrix $\mathcal{C}_\pi$,
\begin{equation}
\left(\begin{array}{c}
  F_\pi \\
  F_{\rm A} \\
  F_{\rm V}
\end{array}\right) =  \mathcal{C}_\pi \left(\begin{array}{c}
  v \\
  f \\
  f
\end{array}\right) = \left(\begin{array}{c}
  v \sqrt{1\!-\!\chi r_3} \\
   \frac{\sqrt{2} M_{\rm A}}{\tilde g}\left(1\!-\!\chi\right) \\
    \frac{\sqrt{2} M_{\rm V}}{\tilde g}
\end{array}\right) ,
\end{equation}
where $\chi=r_3 \tilde g^2v^2/(2M_A^2)$.
As mentioned above, the change of the decay constants originates from a change of the coupling of the involved fields
to the electroweak sector, which mediates the corresponding decays. These circumstances are the same for the five-dimensional setup. There, the condensates are functions of the fifth coordinate $z$. 
We can extract,
\be
\frac{f_\pi^2}{f_\pi^2|_{r_3=0}}
=
\frac{F_\pi^2}{F_\pi^2|_{r_3=0}}
&=&
1-\chi r_3,\\
\frac{F_{\mathcal{A}}}{F_{\mathcal{A}}|_{r_3=0}}
=
\frac{F_{\rm A}}{F_{\rm A}|_{r_3=0}}
&=&
1-\chi,\\
\frac{F_{\mathcal{V}}}{F_{\mathcal{V}}|_{r_3=0}}
=
\frac{F_{\rm V}}{F_{\rm V}|_{r_3=0}}
&=&
1.
\ee
Consequently, the $r_3$ term leads to the following modifications of Eqs.~(\ref{softdecay},~\ref{fpi}),
\be
\sqrt{2}g_5F_\mathcal{A}
&=&
\sqrt{2}\partial^2_z\mathcal{A}(0)(1-\chi)
=\nn&=&
4c_\mathcal{A}(1-\chi)
=\nn&=&
M_\mathcal{A}^2(1-\chi),
\ee
and
\begin{equation}
g^2_5f_\pi^2
=
-\partial^2_z\mathcal{P}(0)(1-\chi r_3)
=
(c_\mathcal{A}-c)(1-\chi r_3) .
\end{equation}
Due to $\chi r_3>0$ and $c>0$ as well as $M_\mathcal{A}^2=4c_\mathcal{A}$ this last relation implies directly $M_\mathcal{A}^2\ge4g_5^2f_\pi^2$.


\section{Benchmarks}

The physical parameters of the model are the masses and decay constants of the lightest vector and axial vector resonances. For technicolour models, the pion decay constant is fixed by the value of the electroweak scale.

In standard holography \cite{Dietrich:2008ni,Polchinski:2000uf,Erlich:2005qh}, the vector resonances do not couple to the condensate and only one parameter remains undetermined, which in \cite{Dietrich:2008ni}, we chose to be the mass of the lightest vector resonance. In particular, the vector resonance had to be always lighter than its axial vector partner.

In the generalised approach \cite{Dietrich:2008up} without $r_3$-term, which does not affect the fifth-dimensional analysis directly, the three additional parameters in the Lagrangian can be summarised into one, i.e., the coupling strength of the vector resonances to the condensate. On the side of the physical parameters, two degrees of freedom remain undetermined, i.e., one more than in the non-generalised approach. The vector resonance can now also be heavier than its axial vector partner. In \cite{Dietrich:2008up}, we chose a benchmark scenario in which we saturated the first Weinberg sum rule by the first pair of resonances. While this is not a necessary requirement, it is still a completely legal point in parameter space and allowed us to compare to findings in non-extradimensional settings \cite{Foadi:2007ue}, which used the same assumption.

\begin{table*}[t]
\begin{tabular}{cccccccccc}
\hline\hline
$N_c$&representation&$d_\mathrm{R}$&$N_f$&$N_f^g$&$\mathcal{G}\rightarrow\mathcal{H}$&
$M_{a,\mathrm{soft}}^\mathrm{min}$&$(F_{a}^\mathrm{min})^{1/2}$\\
\hline
2&fundamental&2&7&6&$SU(7)\times SU(7)\rightarrow SU(7)$&2.2&0.66\\
2&fundamental&2&7&2&$SU(7)\times SU(7)\rightarrow SU(7)$&3.8&1.15\\
2&adjoint&3&2&2&$SU(4)\rightarrow SO(4)$&3.1&1.04\\
3&fundamental&3&11&2&$SU(11)\times SU(11)\rightarrow SU(11)$&3.1&1.04\\
3&2-index-symmetric&6&2&2&$SU(2)\times SU(2)\rightarrow SU(2)$&2.2&0.87\\
3&adjoint&8&2&2&$SU(4)\rightarrow SO(4)$&1.9&0.81\\
4&fundamental&4&15&2&$SU(15)\times SU(15)\rightarrow SU(15)$&2.7&0.97\\
4&2-index-symmetric&10&2&2&$SU(2)\times SU(2)\rightarrow SU(2)$&1.7&0.77\\
4&2-index-antisymmetric&6&8&2&$SU(16)\rightarrow SO(16)$&2.2&0.87\\
4&adjoint&15&2&2&$SU(4)\rightarrow SO(4)$&1.4&0.69\\
5&fundamental&5&19&2&$SU(19)\times SU(19)\rightarrow SU(19)$&2.4&0.91\\
5&2-index-antisymmetric&10&6&2&$SU(6)\times SU(6)\rightarrow SU(6)$&1.7&0.77\\
6&fundamental&6&23&2&$SU(23)\times SU(23)\rightarrow SU(23)$&2.2&0.87\\
\hline
&&&&&&TeV&TeV\\
\hline\hline
\end{tabular}
\caption{
Various walking technicolour models from Tab.~III in
\cite{Dietrich:2006cm}. Minimal (axial) vector meson mass
$M_{a/\rho,\mathrm{soft}}^\mathrm{min}$ (Especially, $r_3=0$.) and square root of the minimal (axial) vector
decay constant $(F_{a/\rho}^\mathrm{min})^{1/2}$
for various walking technicolour
models characterised by the representation of the technicolour gauge group
under which the techniquarks transform and the number of (electroweakly gauged) techniflavours $N_f$ ($N_f^g$).
}
\label{Tab1}
\end{table*}

Taking also into account the $r_3$-term, as was done in the present investigation, 
at the level of the physical parameters, there remains, the relation (\ref{softdecay}) between the vector mass and decay constant. Hence, after the holographic analysis, there are three free physical parameters. This leads to an, in general, non-zero contribution to the oblique $S$-parameter from the first pair of resonances. In the soft-wall setting without $r_3$-term this contribution and the contributions from every pair of resonances was exactly equal to zero \cite{Dietrich:2008up}. Interestingly, without extra-dimensions at tree-level the $S$-parameter is also directly linked to the $r_3$-term. Especially, the contribution is zero without the latter. To the contrary, in the hard-wall setting the contribution to $S$ from the first pair of resonances is, in general, nonzero.  
In order to have concrete benchmark scenarios in the presence of the $r_3$-term, additionally to saturating the first Weinberg sum rule by the first pair of resonances, we, thus, propose to investigate scenarios with fixed contributions from the first pair of resonances to the $S$-parameter. In the non-extradimensional setup \cite{Foadi:2007ue}, e.g., it was set to the one-loop perturbative value $S_\mathrm{pert}=N_f^g d_R/(12\pi)$. Setting it to zero removes the effect from the $r_3$-term altogether. Hence, using $S$ as parameter allows for a smooth interpolation between the different setups. Employing the second Weinberg sum rule for this purpose is less useful because in quasiconformal theories its right-hand side is modified by ultraviolet contributions to the spectral functions and involves a parameter that is not exactly known quantitatively \cite{Appelquist:1998xf}. 

Here, we had set out to construct a model that for quasiconformal theories can produce the linear trajectories for the resonance masses, without the need of having a dilaton. In the dilaton-free case, the discussion from the previous paragraph remains essentially unaltered, despite the fact that the number of unphysical parameters is reduced by one by imposing $c=0$. At the level of the physical parameters, for $c=0$, we find from the above final results,
\be
\sqrt{2}g_5F_\mathcal{V}=4C\lambda_\mathcal{V}=M^2_\mathcal{V},
\label{nodilv}
\ee
\be
\sqrt{2}g_5F_\mathcal{A}
=4C\lambda_\mathcal{A}(1-\chi)
=M^2_\mathcal{A}(1-\chi),
\label{nodila}
\ee
\be
4g^2_5f_\pi^2
=
4C\lambda_\mathcal{A}(1-\chi r_3)
=
M^2_\mathcal{A}(1-\chi r_3) .
\label{fpi2} 
\ee
The relation (\ref{nodilv}) remains unaffected. Imposing a value for $S$,
\be
\frac{F_\mathcal{V}^2}{M_\mathcal{V}^4}-\frac{F_\mathcal{A}^2}{M_\mathcal{A}^4}
=
\frac{S}{4\pi},
\ee
permits us to solve for
$S=S_\mathrm{pert}\chi(2-\chi)/N_g$,
where use has been made of Eq.~(\ref{uv}) and $N_g=N_f^g/2$ is the number of electroweakly gauged techniquark generations.
This leaves $r_3$ free to give $M_\mathcal{A}^2$ any value above $4g_5^2f_\pi^2$ ($\chi r_3\ge0$), as long as $\chi$ (and hence also $S$) is not equal to zero. At the same time, this relation also implies a maximum value for $S\le S_\mathrm{pert}/N_g$. Thus $S$ from the first pair of resonances has always to be smaller than $S_\mathrm{pert}$ per gauged generation to be consistent with the holographic description.


\section{Results}

That for quasiconformal theories the dilaton is not needed to have a constant spacing of the eigenvalues for the squared masses of the (axial-)vector resonances has already been shown in \cite{Dietrich:2008up}. There, however, omission of the dilaton still led to a constrained phenomenology. To the contrary, taking into account the $r_3$ term allows us to remove the dilaton without any impact on the phenomenology. In order to see this, look at the above final results, where $c$ has been set to zero, i.e., Eqs.~(\ref{nodilv}), (\ref{nodila}) and (\ref{fpi2}). 
$\lambda_\mathcal{V}$ appears exclusively in the the expressions for $F_\mathcal{V}$ and $M_\mathcal{V}$. $C$ could be absorbed in a redefinition of $\lambda_\mathcal{V}$ and hence, in this setup, the vectorial and axial sectors are completely decoupled. Further, in this setup, $F_\mathcal{V}$ and $M_\mathcal{V}^2$ always have the fixed ratio (\ref{nodilv}), $\sqrt{2}g_5F_\mathrm{V}=M_\mathrm{V}^2$, where from Eq.~(\ref{uv}), $g_5^2=12\pi^2/d_\mathrm{R}$. (We could not find any parity invariant terms of order four which would change it.)
In the hard-wall model the ratio is $g_5F_\mathcal{V}=1.1328 M_\mathcal{V}^2$ \cite{Dietrich:2008ni} and in a purely four-dimensional approach one has $\tilde g F_\mathcal{V}=2 M_\mathcal{V}^2$. For given $F_\mathcal{V}$ and $M_\mathcal{V}$ this leads consistently to $g_5<\tilde g$.

A redefinition of $C$ in the relations (\ref{nodila}) and (\ref{fpi2}) [and a subsequent redefinition of $\lambda_\mathcal{V}$ in Eq.~(\ref{nodilv})] can absorb the factor of $\lambda_\mathcal{A}$. The new $C$, $\chi$ and $r_3$ are independent except for the condition $\chi r_3\ge0$. Therefore, $F_\mathcal{A}$ and $M_\mathcal{A}$ are decoupled. For $\chi=1$, $F_\mathcal{A}$ may even vanish. The condition $\chi r_3\ge0$ implies $M_\mathcal{A}^2\ge4g_5^2f_\pi^2$, but otherwise $M_\mathcal{A}$ is decoupled from $f_\pi$. (What this means in numbers for the technicolour models listed in \cite{Dietrich:2006cm} is shown in Table \ref{Tab1}.)
Consequently, the omission of the dilaton does not exclude any points from the space of physical parameters $F_\F$ and $M_\F$.

In standard bottom-up holography \cite{Erlich:2005qh,Dietrich:2008ni}, the vector states do not couple to $M_0$ and in the soft-wall model the parameter space is constrained by,
\be
\sqrt{2}g_5F_\mathcal{W}=M^2_\mathcal{W}
\mathrm{~~~and~~~}
M_\mathcal{A}^2=M_\mathcal{V}^2+4g_5^2f_\pi^2,
\ee
which after fixing $f_\pi$ to its physical value leaves, say, $M_\mathcal{V}^2$ to parametrise the rest of the physical quantities.

In generalised bottom-up holography, even after removing the dilaton, only $F_\mathcal{V}$ and $M^2_\mathcal{V}$ remain linked and for $M_\mathcal{A}^2$ lingers a lower bound of $4g_5^2f_\pi^2$ due to the common equation in the axial sector. This corresponds to two additional degrees of freedom.

In a purely four-dimensional treatment, the matching condition (\ref{uv}) of the coupling constant $g_5$ in the ultraviolet is absent which decouples $F_\mathcal{V}$ and $M_\mathcal{V}^2$. Also the lower bound on $M_\mathcal{A}^2$ is absent.


\section*{Acknowledgments}

The authors would like to thank
Francesco Sannino and Joseph Schechter
for helpful and inspiring discussions.
The work of DDD was supported by the Danish Natural Science Research Council.
The work of MJ was supported by the Villum Kann Rasmussen foundation.
The work of CK was supported by the Marie Curie Fellowship under contract MEIF-CT-2006-039211.


\end{document}